\documentclass{elsart}

\usepackage{amssymb}\usepackage{amsmath}
\usepackage[all, knot]{xy}
\xyoption{arc} \xyoption{color}
\usepackage{mathrsfs}
\begin{document}

\begin{frontmatter}

% Title, authors and addresses

% use the thanksref command within \title, \author or \address for footnotes;
% use the corauthref command within \author for corresponding author footnotes;
% use the ead command for the email address,
% and the form \ead[url] for the home page:
% \title{Title\thanksref{label1}}
% \thanks[label1]{}
% \author{Name\corauthref{cor1}\thanksref{label2}}
% \ead{email address}
% \ead[url]{home page}
% \thanks[label2]{}
% \corauth[cor1]{}
% \address{Address\thanksref{label3}}
% \thanks[label3]{}

\title{Perfectly matched layers for the stationary Schr\"{o}dinger equation in a periodic structure}

% use optional labels to link authors explicitly to addresses:
% \author[label1,label2]{}
% \address[label1]{}
% \address[label2]{}

\author{Victor Kalvin\thanksref{AKA}}
\thanks[AKA]{This work was funded by grant number
108898 awarded by the Academy of Finland.} \ead{vkalvin@gmail.com}
\address{Department of Mathematical Information Technology, University of Jyv\"{a}skyl\"{a}, P.O. Box 35 (Agora),
 FIN-40014 University of Jyv\"{a}skyl\"{a}, Finland}

\date{}
\begin{abstract}
We construct a perfectly matched absorbing layer for stationary
Schr\"{o}dinger equation with analytic slowly decaying potential
in a periodic structure. We prove the unique solvability of the
problem with perfectly matched layer of finite length and show
that solution to this problem approximates a solution to the
original problem  with an error that exponentially tends to zero
as the length of perfectly matched layer tends to infinity.

\end{abstract}

\end{frontmatter}

% main text

\section{Introduction }
The Perfectly Matched Layer (PML) method, introduced in [1], is in
common use for a numerical analysis of a wide class of problems.
For some problems the convergence of the method has been proved
mathematically, see e.g. [2--5]. In this paper, we introduce the
PML method for the stationary Schr\"{o}dinger equation in a
``half-plane'' with periodic boundary and Dirichlet boundary
condition. We suppose that the potential $q$ allows an analytic
continuation to a cone on some distance from the boundary and
$q(z)$ uniformly tends to zero as $z$ goes to infinity inside the
cone. We include into consideration potentials decaying at
infinity as slowly  as $z^\nu$, $\nu<0$, or even $1/\ln z$. Since
the potentials are not compactly supported the modal analysis
employed in \cite{ref5} cannot be used here, this leads to
significant difficulties. Using the tools of complex scaling
\cite{ref8,ref9} we construct a PML of infinite length for the
original problem supplied with some generalized radiation
condition. The form of this radiation condition is similar to the
pole condition
 \cite{ref4,ref10}. The generalized radiation condition turns out
to be equivalent to the classical radiation condition in the case
of sufficiently rapid decay of the potential. As an approximation
of a solution satisfying the original problem and the radiation
condition, we take a solution to the problem with PML of finite
length. We prove that the problem with PML of finite length is
uniquely solvable and that the error of the approximation tends to
zero with an exponential rate as the length of PML tends to
infinity. The proof is based on weak statements of problems in
 weighted Sobolev spaces \cite{ref7} and on a modification of the compound
 expansion method \cite{ref6}.

  We consider the
 Dirichlet boundary condition as a boundary condition of the
 original problem and as an artificial boundary condition, however
 one can use the Neumann boundary condition instead. The
 approach is easily extended for this case, the results remain the same.

\section{Statement of the problem}\label{s2}

Let $\mathcal P$ be an upper ``half-plane'' in $\Bbb R^2$ with
smooth $2\pi$-periodic boundary $\partial \mathcal P$ such that
$\mathcal P\subset \{(y,t)\in\Bbb R^2: t>c\}$ and  $\partial
\mathcal P\subset \{(y,t)\in \Bbb R^2: t<0\}$. Let $\mathcal
E=\{(y,t)\in \mathcal P:|y|<\pi\}$ be the periodicity cell of
$\mathcal P$. We set $\Upsilon^\pm=\{(y,t)\in\mathcal
P:y=\pm\pi\}$ and $\Upsilon^0=\partial \mathcal E\setminus
\{\Upsilon^+\cup\Upsilon^-\}$.  As it usually is, the problem in
$\mathcal P$ reduces to a quasi-periodic boundary value problem in
the periodicity cell $\mathcal E$, see Fig.~1.
\begin{figure}[h]
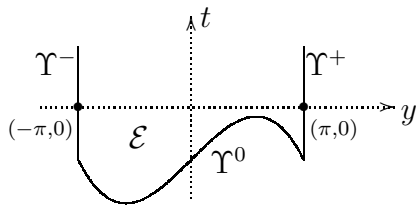

$$\xy  (0,0)*{}="A"; (30,0)*{}="B"; (0,15)*{}="C"; (30,15)="D";
 "A"; "B" **\crv{(10,-20) & (20,20)};
"A";"C" **\dir{-}; "B";"D" **\dir{-};
(-5,4)*{\scriptstyle{(-\pi,0)}}; (34,4)*{\scriptstyle{(\pi,0)}};
(0,7)*{\scriptstyle{\bullet}};(30,7)*{\scriptstyle{\bullet}};
{\ar@{.>} (-5,7)*{}; (42,7)*{}}; (44,6)*{y};
 {\ar@{.>} (15,-5)*{}; (15,19)*{}};
 (17,19)*{t};(8,3)*{\Large{\mathcal E}};
 (-3,13)*{\Upsilon^-};(33,13)*{\Upsilon^+};(20,0)*{\Upsilon^0}
\endxy
$$
\caption{Geometry of the problem.}
\end{figure}

We consider the stationary Schr\"{o}dinger equation
\begin{equation}\label{op1}{(\Delta+k^2+q(y,t))}u(y,t)
= F(y,t),\ (y,t)\in \mathcal E,
\end{equation}
with the quasi-periodicity conditions
\begin{equation}\label{op2}
\partial_y^j u(\pi,t)=e^{2\pi i \alpha} \partial_y^j u(-\pi,t),\  j=0,1,\ (\pm\pi,t)\in
\Upsilon^\pm,
\end{equation}
 and the Dirichlet boundary condition
\begin{equation}\label{op3}
u(y,t) =0,\ (y,t)\in\Upsilon^0.
\end{equation}
Here $\alpha\in[0,1)$, $\partial_y=\partial/{\partial y}$, and the
parameter $k$ is a fixed real number that does not coincide with a
threshold value, i.e. $k^2\neq (n+\alpha)^2$ for all $n\in\Bbb Z$.
Let $K_T^\phi$ denote the closed cone
 $$
 K_T^\phi=\{z\in\Bbb C:
z=T+e^{i\psi}t, 0\leq\psi\leq\phi,t\geq 0\}.
$$
We assume that the potential $q$ in the equation~\eqref{op1}
satisfies the conditions: (i) $q$ is a bounded real-valued
function in ${\mathcal P}$, $q(y,t)=q(y+2\pi,t)$ for all
$(y,t)\in{\mathcal P}$; (ii) for some $T>0$ and $\phi\in(0,\pi/2)$
the potential $q$ can be continued to an analytic in $z$ (and
$2\pi$-periodic in $y$) function $\Bbb R\times
{K^\phi_T}\ni(y,z)\mapsto q(y,z)\in \Bbb C$, which uniformly tends
to zero as $|z|\to +\infty$.

We also make the following assumptions on  the right hand side $F$
of the equation~\eqref{op1}: (i) $F$ is in the space
$L_2^{loc}(\mathcal E)$ of locally square summable functions on
$\mathcal E$; (ii) for some $T>0$ and $\phi\in(0,\pi/2)$ the
function $F$ is an analytic in $z$ function $(-\pi,\pi)\times
K^\phi_T\ni (y,z)\mapsto F(y,z)\in \Bbb C$
 satisfying the uniform in $\psi\in[0,\phi]$ estimate
\begin{equation}\label{tau}
\int_{0}^{+\infty}\int_{-\pi}^\pi \exp(2t\tau\sin\psi )
|F(y,T+e^{i\psi}t)|^2\,dy\,dt\leq Const
\end{equation}
with some $\tau>0$.

\section{Radiation condition and complex scaling}

For all $n\in\Bbb Z$ we set
$\lambda_n^\pm=\mp\sqrt{k^2-(n+\alpha)^2}$, where we take the main
branch of the square root. Let $\mathfrak N=\{n\in\Bbb
Z:|n+\alpha|<|k|\}$. The finite set of points $\{\lambda:
\lambda=\lambda_n^+ \text{ or } \lambda=\lambda_n^-, n\in
\mathfrak N\}$ consists of all points $\lambda_n^\pm$ lying on the
real axis. The remaining points $\lambda_n^\pm$, $n\in\Bbb
Z\setminus\mathfrak N$ are on the imaginary axis. With every
$\lambda_n^\pm$ we associate the function
$w^\pm_n(y,t)=\exp(i\lambda_n^\pm t+i(n+\alpha)y)$.
 The functions
$w_n^\pm$ satisfy the quasi-periodicity conditions \eqref{op2} and
the homogeneous equation \eqref{op1} with $q\equiv 0$. If $n\in
\mathfrak N$ then $w^+_n$ is an incoming wave and $w^-_n$ is an
outgoing wave of the problem \eqref{op1}--\eqref{op3} with
$q\equiv 0$. If $n\in\Bbb Z\setminus\mathfrak N$ then $w^+_n$ is a
growing mode and $w^-_n$ is an evanescent mode of the unperturbed
problem \eqref{op1}--\eqref{op3}. Let $\phi$  be the angle for
which the assumptions of Section~\ref{s2} on the potential $q$ and
the right hand side $F$ are satisfied. We introduce the open cone
$$ \mathcal
K^\phi_\beta=\{\lambda\in\Bbb C:\lambda=i\beta- e^{-i\psi}\xi,
\psi\in(0,\phi), \xi>0\}
$$
with the vertex  $i\beta\in\Bbb C$ and the angle $\phi$,
$\phi\in(0,\pi/2)$.
 Denote by  $H^\ell_\beta
 (\mathcal E)$, $\ell\geq 0$, the weighted   space
 with the norm $\|e_\beta \cdot; H^\ell(\mathcal E)\|$,
 where  $H^\ell(\mathcal E)$ is the Sobolev
 space, and $e_\beta:(y,t)\mapsto\exp \beta t$, $\beta\in\Bbb R$. We say that the parameter $\beta$  is admissible if $\beta\in
(\max\{\Im\lambda_n^+:n\in\Bbb Z\setminus\mathfrak
N\},0)\cap[-\tau\sin\phi,0)$ and the cone $\mathcal K^\phi_\beta$
contains all the points from the set $\{\lambda_n^+:n\in\mathfrak
N\}$, see Fig.~2.

\begin{figure}[h]
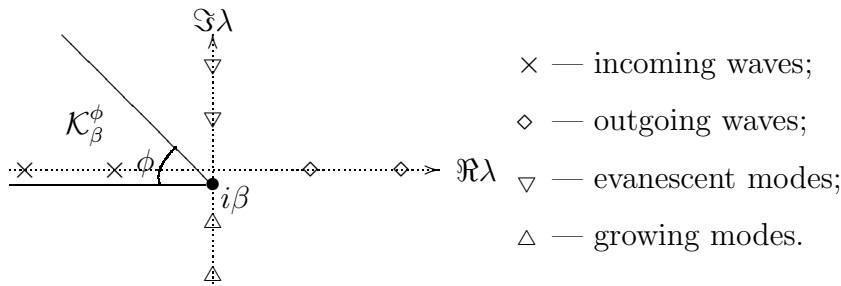

\[ \xy {\ar@{.>}(-7,0); (50,0)};(55,0)*{\Re\lambda};
{\ar@{.>}(20,-15);(20,18)};(20,20)*{\Im\lambda};
(20,-2)*{\bullet}; (-7,-2) **\dir{-};%(20,-3);(50,-3)**\dir{.};
%(45,-3) ;
(23,-4)*{ i\beta}; (20,-2); (0,18) **\dir{-};
(15,3);(13,-2)**\crv{(12,0)};(11,1)*{\phi};
%(10,15)*{\scriptstyle\lambda\in\Bbb C}
(33,0)*{\diamond};(45,0)*{\diamond};
(7,0)*{\times};(-5,0)*{\times};(20,7)*{\scriptstyle\bigtriangledown};
(20,14)*{\scriptstyle\bigtriangledown};(20,-7)*{\scriptstyle\bigtriangleup};
(20,-14)*{\scriptstyle\bigtriangleup}; (3,6)*{\mathcal
K^\phi_\beta}; (82,3)*{
\begin{array}{ll}
     \times &\text{--- incoming waves}; \\
    \diamond & \text{--- outgoing waves}; \\
    {\scriptstyle\bigtriangledown } & \text{--- evanescent modes}; \\
    {\scriptstyle\bigtriangleup} & \text{--- growing modes.} \\
\end{array} }
\endxy\]
\caption{Arrangement of the cone $\mathcal K^\phi_\beta$ for an
admissible $\beta$.}
\end{figure}
\begin{defn} A solution $u$ to
the problem \eqref{op1}--\eqref{op3} satisfies the radiation
condition if for some admissible $\beta$ the solution $u$ is in
the space $H^2_\beta(\mathcal E)$ and  the Fourier-Laplace
transform
$
\hat u (y,\lambda)=\int_{0}^{+\infty} e^{-i\lambda t} u(y,t)\, dt
$
is an analytic in the cone $\mathcal K^\phi_\beta$ function
$\lambda\mapsto \hat u(\cdot,\lambda)$ taking values in the
Sobolev space $H^2(-\pi,\pi)$.
\end{defn}

\begin{thm}\label{t2} Let the assumptions of Section~\ref{s2} be fulfilled.
 (i) If the homogeneous problem \eqref{op1}--\eqref{op3} has no nontrivial
solution in the space $H^2_\gamma(\mathcal E)$ for any $\gamma>0$
then the problem \eqref{op1}--\eqref{op3} has a unique solution
satisfying the radiation condition. (ii) Let a solution $u$ to the
problem \eqref{op1}--\eqref{op3} satisfy the radiation condition
for an admissible $\beta$. Then $u$ satisfies the radiation
condition for every admissible $\beta$.
\end{thm}
We briefly discuss our formulation of radiation condition, for the
details as well as for the proof of Theorem~\ref{t2} we refer to
\cite{ref8}. One can note that our radiation condition looks quite
similar to the pole condition introduced in~\cite{ref4,ref10}  as
an equivalent and universal formulation of the classical radiation
conditions for a wide class of problems. Formally, the only
difference between the pole condition and our radiation condition
is that we require the analyticity of the Fourier-Laplace
transform in a cone instead of the half-plain. Nevertheless, the
classical radiation conditions are not applicable under our
assumptions on the decay of the potential. The introduced
radiation condition should be considered
as a generalization of the classical one. %, the proof of the
%Theorem~\ref{t2} turns out to be laborious and contensive.
In the case of sufficiently rapid decay (say, with an exponential
rate) of the potential $q$ and of the right hand side $F$ at
infinity, our radiation condition is equivalent to the well known
one: a solution satisfies the radiation condition if the principal
term of its asymptotic  at infinity is a linear combination of the
outgoing waves.

Let $\mathcal E^T=\{(y,t)\in\mathcal E: t<T\}$.
 By applying the
complex scaling $t\to T+e^{i\phi}(t-T)$ for $t\geq T$ (complex
change of variables) to the original problem (1--3), we obtain the
problem
\begin{equation}\label{lp1}
\bigl(\Delta+k^2+q(y,t)\bigr)v(y,t)=\mathcal F(y,t),\
(y,t)\in\mathcal E^T,
\end{equation}
\begin{equation}\label{lp2}\begin{aligned}
\bigl(\partial_y^2+e^{-2i\phi}\partial_t^2+q_T^\phi(y,t)+k^2\bigr)v(y,t)=\mathcal
F(y,t),\ (y,t)\in\mathcal E\setminus\overline{\mathcal E^T},
\end{aligned}
\end{equation}
\begin{equation}\label{lp3}
\partial_y^j v|_{\Upsilon^+}=e^{2\pi i \alpha} \partial_y^j v|_{\Upsilon^-},\ \ j=0,1,
\end{equation}
\begin{equation}\label{lp4}
v =0\ \text{on\ }\Upsilon^0
\end{equation}
\begin{equation}\label{lp5}
 \partial_t^j v(y,T-)=e^{-i\phi j}\partial_t^j v(y,T+),\ j=0,1,
|y|\leq\pi,
\end{equation}
where the equation \eqref{lp5} is a jump condition, $\partial_t^j
v(y,T-)$ and $\partial_t^j v(y,T+)$ denote the limits of
$\partial_t^j v(y,t)$ as $t$ tends to  $T$ from the left and from
the
 right side correspondingly.
The potential $q^\phi_T$ in \eqref{lp2} is defined by the equality
\begin{equation}\label{q}
q^\phi_T(y,t)=q(y,T+e^{i\phi}(t-T)), \quad (y,t)\in\mathcal
E\setminus\overline{\mathcal E^T},
\end{equation}
the right hand side $\mathcal F$ in \eqref{lp1},\eqref{lp2} is
given by
\begin{equation}\label{F}
\mathcal F(y,t)=\left\{%
\begin{array}{ll}
   F(y,t), & (y,t)\in\mathcal E^T, \\
    F(y,T+e^{i\phi}(t-T)),& (y,t)\in\mathcal
E\setminus\overline{\mathcal E^T}. \\
\end{array}%
\right.
\end{equation}
The problem~\eqref{lp1}--\eqref{lp5}
 is elliptic for $\phi\in(0,\pi/2)$.

Let $H_\gamma^{1,\alpha}(\mathcal E)$ denote the closed subspace
in $H_\gamma^1(\mathcal E)$ of all functions satisfying the
quasi-periodicity condition $u|_{\Upsilon^+}=e^{2\pi i\alpha}
u|_{\Upsilon^-}$. By
$\lefteqn{\stackrel{\circ}{\phantom{\,\,>}}}H_{\gamma}^{1,\alpha}(\mathcal
E)$ we denote the space of all functions $u\in
H_\gamma^{1,\alpha}(\mathcal E)$ such that $u=0$ on $\Upsilon^0$.
Then we introduce the space $H_\gamma^{-1,\alpha}(\mathcal E)$ as
the dual space of
 $\lefteqn{\stackrel{\circ}{\phantom{\,\,>}}}H_{-\gamma}^{1,\alpha}(\mathcal
E)$ endowed with the natural norm
\begin{equation}\label{norm}
\|\mathcal F;H_\gamma^{-1,\alpha}(\mathcal E)\|=\sup\{|(\mathcal
F,w)_{\mathcal E}|: w\in
\lefteqn{\stackrel{\circ}{\phantom{\,\,>}}}H_{-\gamma}^{1,\alpha}(\mathcal
E),
\|w;\lefteqn{\stackrel{\circ}{\phantom{\,\,>}}}H_{-\gamma}^{1,\alpha}(\mathcal
E)\|=1\},
\end{equation}
 where $(\cdot,\cdot)_{\mathcal E}$ is the
extension of the inner product in $L_2(\mathcal E)$ to the pairs
$(\mathcal F,w)\in H_\gamma^{-1,\alpha}(\mathcal E)\times
\lefteqn{\stackrel{\circ}{\phantom{\,\,>}}}H_{-\gamma}^{1,\alpha}(\mathcal
E)$. Let us note that if the right hand side $F$ of the
equation~\eqref{op1} satisfies the assumption (ii) from
Section~\ref{s2} for some $\psi$, $T=T_0$, and $\tau=\tau_0$ then
$F$ satisfies the uniform in $\psi\in[0,\varphi]$ estimate
\eqref{tau} for every $T>T_0$, $\tau\leq\tau_0$, and
$\varphi\leq\phi$, see \cite{ref9}. Without loss of generality we
can assume that the potential $q$ and the right hand side $F$
satisfy the assumptions of Section~\ref{s2} for some $\tau>0$,
$\phi\in(0,\pi/2)$, and for all sufficiently large positive $T$.
Due to~\eqref{tau} and~\eqref{F} we have $\mathcal F\in
H^0_{\gamma}(\mathcal E)$ for all $\gamma\leq \tau\sin\phi$, it is
clear that $\|\mathcal F;H_\gamma^{-1,\alpha}(\mathcal
E)\|\leq\|\mathcal F; H^0_\gamma(\mathcal E)\|$.

 Consider the variational statement of the problem
\eqref{lp1}--\eqref{lp5}: {\it find a function $v\in
\lefteqn{\stackrel{\circ}{\phantom{\,\,>}}}H_{\gamma}^{1,\alpha}(\mathcal
E)$ satisfying the equation
\begin{equation*}
\begin{aligned}
-e^{-i\phi}\int_{\mathcal E^T}\bigl(\partial_y
v\cdot\partial_y\bar w+&\partial _t v\cdot\partial_t \bar
w-(q(y,t)+k^2) v\cdot \bar
w\bigr)\,dy\,dt\\
-\int_{\mathcal E\setminus\overline{\mathcal E^T}}\bigl(\partial_y
v\cdot\partial_y\bar w+&e^{-2i\phi}\partial _t v\cdot\partial_t
\bar w-(q_T^\phi(y,t)+k^2) v\cdot \bar
w\bigr)\,dy\,dt\\&=e^{-i\phi}(\mathcal F,w)_{\mathcal
E^T}+(\mathcal F,w)_{\mathcal E\setminus\overline{\mathcal
E^T}}\quad\forall w\in
\lefteqn{\stackrel{\circ}{\phantom{\,\,>}}}H^{1,\alpha}_{-\gamma}(\mathcal
E).
\end{aligned}
\end{equation*}}
The variational form of the problem \eqref{lp1}--\eqref{lp5}
generates the linear continuous operator
\begin{equation}\label{mA}
\lefteqn{\stackrel{\circ}{\phantom{\,\,>}}}H_{\gamma}^{1,\alpha}(\mathcal
E)\ni v\mapsto \mathcal A_\gamma v\in\mathcal F\in
H_\gamma^{-1,\alpha}(\mathcal E).
\end{equation}
%the operator \eqref{mA} is Fredholm (i.e. the range of the
%operator is closed,  kernel and  cokernel are finite-dimensional)
%if and only if the line $\{\lambda\in\Bbb C:\Im\lambda=\gamma\}$
%is free of the points $e^{i\phi}\lambda_n^\pm$, $n\in\Bbb Z$.

The proof of the following proposition can be found in
\cite{ref8}.
\begin{prop}\label{p3}(i) Let the potential $q$ satisfy the assumptions of
Section~\ref{s2}, and let $T$ be a sufficiently large positive
number. We define the potential $q^T_\phi$ for the equation
\eqref{lp2} by the equality \eqref{q}. If the homogeneous problem
\eqref{op1}--\eqref{op3} has no nontrivial solution in the space
$H^2_\gamma(\mathcal E)$ for any $\gamma>0$ then the operator
\eqref{mA} of the problem \eqref{lp1}--\eqref{lp5} yields an
isomorphism if and only if $|\gamma|<\min_{n\in\Bbb
Z}\{\Im(e^{i\phi}\lambda_n^-)\}$.

 (ii) Assume that the potential $q$ and the right hand side $F$ of
 the problem~\eqref{op1}--\eqref{op3} satisfy the assumptions of
Section~\ref{s2}. Let $T$ be a sufficiently large positive number.
We define the potential $q^T_\phi$ and the right hand side
$\mathcal F$ of the problem~\eqref{lp1}--\eqref{lp5} by the
equalities \eqref{q}, \eqref{F}. Let $u$ be a (unique) solution to
the problem \eqref{op1}--\eqref{op3} satisfying the radiation
conditions. Then a (unique) solution
$v\in\lefteqn{\stackrel{\circ}{\phantom{\,\,>}}}H_{0}^{1,\alpha}(\mathcal
E)$ to the problem \eqref{lp1}--\eqref{lp5} is the analytic
continuation of $u$ in the sense that  $v=u$ on $\mathcal E^T$ and
$v(y,t)=u(y,T+e^{i\phi}(t-T))$ for $(y,t)\in\mathcal
E\setminus\mathcal E^T$.

\end{prop}

\section{PML method. Rate of convergence and error estimate}
 We search for an approximation in a
domain $\mathcal E^L$, $0<L<T$, of a solution $u$ to the problem
\eqref{op1}--\eqref{op3} subjected to the radiation condition.
Since a solution $v\in
\lefteqn{\stackrel{\circ}{\phantom{\,\,>}}}H^{1,\alpha}_0(\mathcal
E)$ to the problem \eqref{lp1}--\eqref{lp5} and $u$ are coincident
on $\mathcal E^T$ (see Proposition~\ref{p3}, ii), one can search
for an approximation of $v$ instead of an approximation of $u$.
The advantage is that $v$ is in the space $H^1_\gamma(\mathcal
E)$, $0<\gamma<\min\{\Im(e^{i\phi}\lambda_n^-)\}$, of functions
``exponentially decaying'' at infinity, while $u\notin
H^1(\mathcal E)$. It is clear that $v$ has these properties
because of the {\it perfectly matched} equation \eqref{lp2}. In
other words, the equation \eqref{lp2} describes a PML
 of infinite length.

We truncate the domain $\mathcal E$ at a finite distance $R>T$.
%, see Fig.~3.
By $\Upsilon^R$ we denote the boundary of truncation,
$\Upsilon^R=\partial \mathcal E^R\setminus\partial \mathcal E$.
Let us also set $\Upsilon^{\pm,R}=\{(y,t)\in\Upsilon^\pm: t<R\}$.
%\begin{figure}[h]
%\[\xy  (0,0)*{}="A"; (30,0)*{}="B"; (0,25)*{}="C"; (30,25)="D";
% "A"; "B" **\crv{(10,-20) & (20,20)};
%"A";"C" **\dir{-}; "B";"D"
%**\dir{-};(37,25)*{(\pi,R)};(30,25)*{\scriptstyle{\bullet}};(15,27)*{\Upsilon^R};
%%(-1,-3)*{-\pi};
%"C";"D"**\dir{-};
% %(31,-3)*{\pi};
% % {\ar@{.>}(-5,0)*{}; (35,0)*{}}; (37,-1)*{y};
% %{\ar@{.>} (15,-5)*{}; (15,30)*{}};(17,30)*{t};
% (10,5)*{{\mathcal E^T}};
% (15,17)*{\mathcal E^R\setminus\mathcal E^T};
% (-5,17)*{\Upsilon^{-,R}};(19,0)*{\Upsilon^0};
% (0,10);(30,10)*{\scriptstyle{\bullet}} **\dir{-};(37,10)*{(\pi,T)}
%% (40,40)*{
%%(\Delta+k^2+q)v=\mathcal F,\text{ in } \mathcal E^T,};
%\endxy
%\]
%\caption{Truncated domain $\mathcal E^R$.}
%\end{figure}
With the aim of approximating $v$ by a solution $v^{R}$ to some
problem in the bounded domain $\mathcal E^R$, we introduce the
problem
\begin{equation}\label{e1} \bigl(\Delta+k^2+q\bigr)v^{R}(y,t)=\mathcal F(y,t),\quad (y,t)\in\mathcal E^T,
\end{equation}
\begin{equation}\label{e2}\begin{aligned}
\bigl(\partial_y^2+e^{-2i\phi}\partial_t^2+q_T^\phi(y,t)+k^2\bigr)v^{R}(y,t)=\mathcal
F(y,t),\ (y,t)\in\mathcal E^R\setminus\overline{\mathcal E^T},
\end{aligned}
\end{equation}
\begin{equation}\label{e3}
\partial_y^j v^{R}|_{\Upsilon^{+,R}}=e^{2\pi i \alpha} \partial_y^j v^{R}|_{\Upsilon^{-,R}},\ \ j=0,1,
\end{equation}
\begin{equation}\label{e4}
v^{R} =0\ \ \ \ \text{{\it on}\ }\Upsilon^0,
\end{equation}
\begin{equation}\label{e5}
 \partial_t^j v^{R}(y,T-)=e^{-i\phi j}\partial_t^j v^{R}(y,T+),\ j=0,1,
\end{equation}
\begin{equation}\label{e6}
v^{R}=\mathcal G \ \ \text{{\it on}\ }\Upsilon^R,
\end{equation}
where as an artificial boundary condition on $\Upsilon^R$ we take
the Dirichlet boundary condition. The equation \eqref{e2}
describes a PML of the finite length $R-T$.

Let
 $\tilde H^{1,\alpha}(\mathcal E^R)$ denote the closed subspace in
$H^1(\mathcal E^R)$ of all functions satisfying the
quasi-periodicity condition $v|_{\Upsilon^{+,R}}=e^{2\pi i\alpha}
v|_{\Upsilon^{-,R}}$ and the boundary condition
$v|_{\Upsilon_0}=0$. By
$\lefteqn{\stackrel{\circ}{\phantom{\,\,>}}}H^{1,\alpha}(\mathcal
E^R)$ we denote the space of all functions $v\in
H^{1,\alpha}(\mathcal E^R)$ such that $v=0$ on
$\Upsilon^0\cup\Upsilon^R$. Then we introduce the space
$H^{-1,\alpha}(\mathcal E^R)$ as the dual space of
$\lefteqn{\stackrel{\circ}{\phantom{\,\,>}}}H^{1,\alpha}(\mathcal
E^R)$. Consider the variational statement of the problem
\eqref{e1}--\eqref{e6}: {\it find a function $v^R\in \tilde
H^{1,\alpha}(\mathcal E^R)$ satisfying the equation
\begin{equation*}
\begin{aligned}
-e^{-i\phi}\int_{\mathcal E^T}\bigl(\partial_y
v^R\cdot\partial_y\bar w+&\partial _t v^R\cdot\partial_t \bar
w-(q(y,t)+k^2) v^R\cdot \bar
w\bigr)\,dy\,dt\\
-\int_{\mathcal E^R\setminus\overline{\mathcal
E^T}}\bigl(\partial_y v^R\cdot\partial_y\bar
w+&e^{-2i\phi}\partial _t v^R\cdot\partial_t \bar
w-(q_T^\phi(y,t)+k^2) v^R\cdot \bar
w\bigr)\,dy\,dt\\&=e^{-i\phi}(\mathcal F,w)_{\mathcal
E^T}+(\mathcal F,w)_{\mathcal E^R\setminus\overline{\mathcal
E^T}}\quad\forall w\in
\lefteqn{\stackrel{\circ}{\phantom{\,\,>}}}H^{1,\alpha}(\mathcal
E^R)
\end{aligned}
\end{equation*}
and the boundary condition $v^R=\mathcal G$ on $\Upsilon^R$.} The
variational statement generates the linear continuous operator
\begin{equation}\label{A^R}
\tilde H^{1,\alpha}(\mathcal E^R)\ni v^R\mapsto \mathcal
A^Rv^R=\{\mathcal F,\mathcal G\}\in H^{-1,\alpha}(\mathcal
E^R)\times H^{1/2,\alpha}(\Upsilon^R),
\end{equation}
where  $H^{1/2,\alpha}(\Upsilon^R)$ is the space of traces on
 $\Upsilon^R$ of
the functions from $\tilde H^{1,\alpha}(\mathcal E^R)$.

\begin{prop}\label{p5} Let $T$ be a sufficiently large positive number and $\phi\in(0,\pi/2)$. Assume
that for all $\gamma>0$ there is no nontrivial solution to the
original homogeneous problem \eqref{op1}--\eqref{op3} in the space
$H^2_\gamma(\mathcal E)$. Then there exists $R_0>T$ such that for
all $R>R_0$ the problem \eqref{e1}--\eqref{e6} with  right hand
side $\{\mathcal F,\mathcal G\}\in H^{-1,\alpha}(\mathcal
E^R)\times H^{1/2,\alpha}(\Upsilon^R)$   admits a unique
variational  solution $v^{R}\in \tilde H^{1,\alpha}(\mathcal
E^R)$. The estimate
\begin{equation}\label{est1}
\|v^{R};H^{1,\alpha}(\mathcal E^R)\|\leq C (\|\mathcal
F;H^{-1,\alpha}(\mathcal E^R)\|+\|\mathcal
G;H^{1/2,\alpha}(\Upsilon^R)\|)
\end{equation}
is valid, where the constant $C$ does not depend on $R>R_0$.
\end{prop}
\begin{pf} The proof is carried out by a modification of the compound expansion
method \cite{ref6}. In other words, we find an approximate
solution to the problem \eqref{e1}--\eqref{e6} compounded of
solutions to first and second limit problems. As the first limit
problem we take the scaled problem \eqref{lp1}--\eqref{lp5}. The
second limit problem is the elliptic problem with constant
coefficients
\begin{equation}\label{1}
\begin{aligned}
\bigl(\partial_y^2+e^{-2i\phi}\partial_t^2+k^2\bigr)\mathsf
U_2(y,t)=\mathsf F_2(y,t),\
(y,t)\in\Pi^R_-,\\
\partial_y^j
\mathsf U_2(\pi,t)=e^{2\pi i \alpha} \partial_y^j \mathsf
U_2(-\pi,t),\ \ j=0,1,\
t<R,\\
\mathsf U_2=\mathsf G_2\  \text{ on } \Upsilon^R,
\end{aligned}
\end{equation}
where $\Pi^R_-=\{(y,t):y\in(-\pi,\pi), t<R\}$.

Let us define the functional spaces for the problem \eqref{1}. By
$H^{1}_\gamma(\Pi^R_-)$ we denote the weighted Sobolev space with
the norm $ \|u;H^{1}_\gamma(\Pi^R_-)\|=e^{-\gamma R}\|e_\gamma u;
H^1(\Pi^R_-)\|$. The space $H^{1,\alpha}_\gamma(\Pi^R_-)$ is the
closed subspace in
 $H^{1}_\gamma(\Pi^R_-)$ of all elements satisfying
$u(\pi,t)=e^{2\pi i \alpha} u(-\pi,t)$ for $t<R$. The space
$\lefteqn{\stackrel{\circ}{\phantom{\,\,>}}}H^{1,\alpha}_{-\gamma}(\Pi^R_-)$
consists of all elements $u\in H^{1,\alpha}_\gamma(\Pi^R_-)$
having the traces $u|_{\Upsilon^R}=0$. We set
$H^{-1,\alpha}_\gamma(\Pi^R_-)=(\lefteqn{\stackrel{\circ}{\phantom{\,\,>}}}H^{1,\alpha}_{-\gamma}(\Pi^R_-))^*$,
the space $H^{-1,\alpha}_\gamma(\Pi^R_-)$ is provided with the
natural norm, cf. \eqref{norm}. Consider the variational statement
of the problem \eqref{1}: {\it find a function $\mathsf U_2\in
H^{1,\alpha}_\gamma(\Pi^R_-)$ which satisfies the equation
$$
\int_{\Pi^R_-}(-\partial_y \mathsf U_2\cdot\partial_y\bar
w-e^{-2i\phi}\partial_t\mathsf
U_2\cdot\partial_t\bar{w}+k^2\mathsf U_2\cdot\bar w)
\,dy\,dt=(\mathsf F_2,w)_{\Pi^R_-}\quad\forall w\in
\lefteqn{\stackrel{\circ}{\phantom{\,\,>}}}H^{1,\alpha}_{-\gamma}(\Pi^R_-)
$$
and the boundary condition $ \mathsf U_2=\mathsf G_2$ on
$\Upsilon^R$.} To the variational form of the problem \eqref{1}
there corresponds the linear continuous operator
\begin{equation}\label{A}
H^{1,\alpha}_{-\gamma}(\Pi^R_-)\ni\mathsf U_2\mapsto
A_{-\gamma}\mathsf U_2=\{\mathsf F_2,\mathsf G_2\}\in
H^{-1,\alpha}_{-\gamma}(\Pi^R_-)\times H^{1/2,\alpha}(\Upsilon^R).
\end{equation}
As is well known \cite{ref7}, the operator $A_{-\gamma}$ is
Fredholm (i.e. the range of the operator is closed,  kernel and
cokernel are finite-dimensional) if and only if there are no
numbers $e^{i\phi}\lambda_n^\pm$, $n\in\Bbb Z$, on the line
$\{\lambda\in\Bbb C: \Im\lambda=\gamma\}$. Suppose that
$\gamma\in[0,\min_n\{\Im(e^{i\phi}\lambda_n^-)\})$. Then
$e^{i\phi}\lambda_n^\pm\notin \{\lambda\in\Bbb C:
\Im\lambda=\gamma\}$, $n\in\Bbb  Z$.
 The solutions to the homogeneous problem
\eqref{1} are easily found in an explicit form, one can see that
they do not belong to the space $H^1_{-\gamma}(\Pi^R_-)$.
Analogously, we consider the formally adjoint to~\eqref{1}
homogeneous problem, and check that it has no solution in the
space $H^1_{\gamma}(\Pi^R_-)$. Therefore, if
$\gamma\in[0,\min_n\{\Im(e^{i\phi}\lambda_n^-)\})$  then the
operator \eqref{A} implements an isomorphism,  a variational
 solution $\mathsf U_2\in H^{1,\alpha}_{-\gamma}(\Pi^R_-)$ of the second limit problem~\eqref{1}
with
 right hand side $\{\mathsf F_2,\mathsf G_2\}\in H^{-1,\alpha}_{-\gamma}(\Pi^R_-)\times H^{1/2,\alpha}(\Upsilon^R)$
  satisfies the estimate
\begin{equation}\label{es3}
 \|\mathsf U_2;H^{1,\alpha}_{-\gamma}(\Pi^R_-)\|\leq C(\|\mathsf
F_2;H^{-1,\alpha}_{-\gamma} (\Pi^R_-)\|+\|\mathsf
G_2;H^{1/2,\alpha}(\Upsilon^R)\|).
\end{equation}
The constant $C$ in~\eqref{es3} is independent of $R$ because the
problem~\eqref{1} reduces to the same problem with $R=0$ by the
shift $t\mapsto t+R$, and the norms in \eqref{es3} are invariant
with respect to $R$, e.g. $\|\mathsf
U_2;H^1_\gamma(\Pi_-^R)\|=\|\mathsf
U_2(\cdot+R);H^1_\gamma(\Pi_-^0)\|$.

Now we are in position to construct the approximate solution.
 Let $\chi$ be a
smooth cut-off function on the real line, $\chi(t)=0$ for $t>1$
and $\chi(t)=1$ for $t<-1$. We denote $\chi_{R/2}(t)=\chi(t-R/2)$,
$t\in\Bbb R$. For a sufficiently large $R>0$ we set $ \mathsf
F_1=\chi_{R/2}\mathcal F$ and $\mathsf F_2=(1-\chi_{R/2})\mathcal
F$, where $(y,t)\in\mathcal E^R$ and $\mathcal F\in
H^{-1,\alpha}(\mathcal E^R)$ is the right hand side of the problem
\eqref{e1}--\eqref{e6}. We extend the functional $\mathsf F_1$
(the functional $\mathsf F_2$) by zero to all $t\geq R$ (to all
$t\leq 0$). It is clear that $\mathcal F=\mathsf F_1+\mathsf F_2$,
for  all $\gamma\in[0,\min_n\{\Im(e^{i\phi}\lambda_n^-)\})$ we
have
\begin{equation}\label{es1}
\|\mathsf F_1; H^{-1,\alpha}_\gamma (\mathcal E)\|\leq e^{\gamma
(R/2+1)}\|\mathcal F;H^{-1,\alpha}(\mathcal E^R)\|,
\end{equation}
\begin{equation}\label{es2}
\|\mathsf F_2; H^{-1,\alpha}_{-\gamma} (\Pi^R_-)\|\leq e^{\gamma
(R/2+1)}\|\mathcal F;H^{-1,\alpha}(\mathcal E^R)\|.
\end{equation}
% Indeed, e.g. the estimate \eqref{es1}
%follows from the evident inequalities
%$$
%\frac {|(\mathsf F_1, w)_{\mathcal
%E}|}{\|w;\lefteqn{\stackrel{\circ}{\phantom{\,\,>}}}H_{-\gamma}^{1,\alpha}(\mathcal
%E)\|}\leq \frac {|(\chi_{R/2}\mathcal F,\eta_{R/2} w)_{\mathcal
%E}|}{\|\eta_{R/2}w;\lefteqn{\stackrel{\circ}{\phantom{\,\,>}}}H_{-\gamma}^{1,\alpha}(\mathcal
%E)\|}\leq e^{\gamma(R/2+1)} \frac {|(\mathcal F,\eta_{R/2}
%w)_{\mathcal
%E}|}{\|\eta_{R/2}w;\lefteqn{\stackrel{\circ}{\phantom{\,\,>}}}H^{1,\alpha}(\mathcal
%E^R)\|},
%$$
%where $\eta_{R/2}$ stands for a smooth cut-off function on $\Bbb
%R$ such that $\chi_{R/2}\bar\eta_{R/2}=\chi_{R/2}$ and
%$\eta_{R/2}(t)=0$ for $t\geq R/2+1$.

  Let $\mathsf U_1\in
\lefteqn{\stackrel{\circ}{\phantom{\,\,>}}}H^{1,\alpha}_\gamma(\mathcal
E)$ be a (unique) solution to the first limit problem
\eqref{lp1}--\eqref{lp5} with the right hand side $\mathsf F_1$,
and let $\mathsf U_2\in H^{1,\alpha}_\gamma(\Pi^R_-)$ be a
(unique) solution to the second limit problem \eqref{1} with the
right hand side $\mathsf F_2$ and $\mathsf G_2\equiv\mathcal G$,
where $\mathcal G$ is the same as in (18), $\mathcal G\in
H^{1/2,\alpha}(\Upsilon^R)$. Due to the first assertion of
Proposition~\ref{p3}, the estimate
\begin{equation}\label{es0}
\|\mathsf U_1;H^{1,\alpha}_\gamma(\mathcal E)\|\leq C\|\mathsf
F_1;H^{-1,\alpha}_{\gamma}(\mathcal E)\|
\end{equation}
is valid.
 We define
the approximate variational solution $Y\in \tilde
H^{1,\alpha}(\mathcal E^R)$ to the problem \eqref{e1}--\eqref{e6}
by the equality
$$
Y(y,t)=\chi(t-2R/3) \mathsf U_1(y,t)+(1-\chi(t-R/3))\mathsf
U_2(y,t),\  (y,t)\in\mathcal E^R.
$$
By setting $\gamma=0$ in the estimates \eqref{es3}, \eqref{es1},
\eqref{es2}, and \eqref{es0},   we derive
\begin{equation}\label{es9}
\|Y; H^{1,\alpha}(\mathcal E^R)\|\leq Const (\|\mathcal
F;H^{-1,\alpha}(\mathcal E^R)\|+\|\mathcal
G;H^{1/2,\alpha}(\Upsilon^R)\|)
\end{equation}
with some constant independent of  $\mathcal F$, $\mathcal G$, and
$R$.
%It can be easily seen  that $\mathcal A^RY=\{\tilde{\mathcal
%F},\mathcal G\}$ with the same $\mathcal G$ as in $\eqref{e6}$ and
%some new right hand side $\tilde {\mathcal F}\in
%H^{-1,\alpha}(\mathcal E^R)$

  On the next step we estimate the
discrepancy that  $Y$ leaves in the right hand side of the problem
\eqref{e1}--\eqref{e6}, in other words, we estimate the value
$$
\|\mathcal A^R Y-\{\mathcal F,\mathcal G\}; H^{-1,\alpha}(\mathcal
E^R)\times H^{1/2,\alpha}(\Upsilon^R)\|;$$ here $\mathcal A^R$ is
the operator \eqref{A^R}.  Recall that by $\mathcal A_\gamma$ and
$A_{-\gamma}$ we denote the operators \eqref{mA} and \eqref{A} of
the first and second limit problems. It is clear that the mappings
\begin{equation*}
\begin{aligned}
\lefteqn{\stackrel{\circ}{\phantom{\,\,>}}}H^{1,\alpha}_0(\mathcal
E)\ni \mathsf U_1 &\mapsto \mathcal A_0\chi_{2/3R}\mathsf U_1\in
H^{-1,\alpha}(\mathcal E^R), \\
H^{1,\alpha}_{0}(\Pi^R_-)\ni \mathsf U_2 & \mapsto
A_0(1-\chi_{R/3})\mathsf U_2\in H^{-1,\alpha}(\mathcal E^R)\times
H^{1/2,\alpha}(\Upsilon^R)
\end{aligned}
\end{equation*}
are continuous. We have
\begin{equation}\label{raz}
\mathcal A^RY=
 \{\mathcal A_0\chi_{2R/3}\mathsf U_1,0\}+\{q^\phi_T(1-\chi_{R/3})\mathsf U_2,0\}+A_{0}(1-\chi_{R/3})\mathsf
 U_2=\{\tilde{\mathcal F},\mathcal G\},
\end{equation}
where we assume that the function $q^\phi_T(1-\chi_{R/3})$ is
extended to $\overline{\mathcal E^R}$ by zero. From \eqref{raz} it
follows that
\begin{equation}\label{dva}
\begin{aligned}
\{\tilde{\mathcal F},\mathcal G\}-\{\mathcal F,\mathcal G\}
=\{\mathsf F_1+[\mathcal A_0,\chi_{2R/3}]\mathsf
U_1,0\}+\{q^\phi_T(1-\chi_{R/3})\mathsf U_2,0\}\\ +\{\mathsf
F_2,\mathcal G\}-[A_0,\chi_{R/3}]\mathsf U_2-\{\mathcal F,\mathcal
G\}\\=\{[\mathcal A_0,\chi_{2R/3}]\mathsf
U_1,0\}+\{q^\phi_T(1-\chi_{R/3})\mathsf U_2,0\}
-[A_0,\chi_{R/3}]\mathsf U_2;
\end{aligned}
\end{equation}
here $[a,b]=ab-ba$. The term $[\mathcal A_0,\chi_{2R/3}]\mathsf
U_1$ is equal to zero outside of the set $\{(y,t):y\in[-\pi,\pi],
t\in[2R/3-1, 2R/3+1 ]\}$. We get
\begin{equation}\label{es6}
\begin{aligned}
\|[\mathcal A_0,\chi_{2R/3}]\mathsf U_1;H^{-1,\alpha}(\mathcal
E^R)\|\leq C e^{-2\gamma R/3}\|\mathsf
U_1;\lefteqn{\stackrel{\circ}{\phantom{\,\,>}}}
H^{1,\alpha}_\gamma(\mathcal E)\|\\\leq C e^{-\gamma
R/6}\|\mathcal F;H^{-1,\alpha}(\mathcal E^R)\|,
\end{aligned}
\end{equation}
where  the last estimate is a consequence of the estimates
\eqref{es1} and \eqref{es0}, the constant $C$ does not depend on
$R$. A similar reasoning together with \eqref{es3} and \eqref{es2}
leads to the estimates
\begin{equation}\label{es8}
\begin{aligned}
\|([ A_0, \chi_{R/3}]\mathsf U_2)_1; H^{-1,\alpha}(\mathcal
E^R)\|\leq C e^{ -2\gamma R/3}\|\mathsf U_2;
H^{1,\alpha}_{-\gamma}(\Pi^R_-)\|\\ \leq C e^{ -2\gamma
R/3}(\|\mathsf F_2; H^{-1,\alpha}_{-\gamma}(\Pi^R_-)\|+\|\mathcal
G; H^{1/2,\alpha}(\Upsilon^R)\|)
\\
\leq C e^{-\gamma R/6}(\|\mathcal F;H^{-1,\alpha}(\mathcal
E^R)\|+\|\mathcal G; H^{1/2,\alpha}(\Upsilon^R)\|)
\end{aligned}
\end{equation}
for the first component
 of the pair $[ A_0,
\chi_{R/3}]\mathsf U_2=\{([ A_0, \chi_{R/3}]\mathsf U_2)_1,0\}$.
 At last, due to our assumptions
on the potential $q$ (see Section~\ref{s2}), we have
\begin{equation}\label{es4}
\|q^\phi_T(1-\chi_{R/3})\mathsf U_2;H^{-1,\alpha}(\mathcal
E^R)\|\leq c(R)\|\mathsf U_2;H^{1,\alpha}_0(\Pi^R_-)\|,
\end{equation}
where $c(R)$ tends to zero as $R\to+\infty$. From \eqref{es4} and
the estimates  \eqref{es3}, \eqref{es2} with $\gamma=0$,
 we see that
\begin{equation}\label{es5}
\|q^\phi_T(1-\chi_{R/3})\mathsf U_2;H^{-1,\alpha}(\mathcal
E^R)\|\leq C c(R)(\|\mathcal F;H^{-1,\alpha}(\mathcal
E^R)\|+\|\mathcal G; H^{1/2,\alpha}(\Upsilon^R)\|).
\end{equation}
Taking into account the equalities \eqref{dva} and the estimates
\eqref{es6}, \eqref{es8}, and \eqref{es5},  we arrive at the
estimate
\begin{equation}\label{es7}
\|\tilde {\mathcal F}-\mathcal F; H^{-1,\alpha}(\mathcal
E^R)\|\leq C(R)(\|\mathcal F; H^{-1,\alpha}(\mathcal
E^R)\|+\|\mathcal G; H^{1/2,\alpha}(\Upsilon^R)\|),
\end{equation}
where $C(R)$ does not depend on $\{\mathcal F,\mathcal G\}$ and
$C(R)\to 0$ as $R\to +\infty$.

We first assume that $\mathcal G=0$. In this case we have
$\tilde{\mathcal F}=\mathcal F+\mathfrak O(R) \mathcal F$ with
some operator $\mathfrak O(R)$ in $H^{-1,\alpha}(\mathcal E^R)$,
whose norm tends to zero as $R\to+\infty$. Hence for a
sufficiently large $R_0$ and for all $R>R_0$ we have
$|\mspace{-2mu}|\mspace{-2mu}|\mathfrak
O(R)|\mspace{-2mu}|\mspace{-2mu}|_R\leq
|\mspace{-2mu}|\mspace{-2mu}|\mathfrak
O(R_0)|\mspace{-2mu}|\mspace{-2mu}|_{R_0}<1$, where
$|\mspace{-2mu}|\mspace{-2mu}|\cdot|\mspace{-2mu}|\mspace{-2mu}|_R$
stands for the operator norm in $H^{-1,\alpha}(\mathcal E^R)$.
There exists the operator
 %\begin{equation}\label{OPER}
 $(I+\mathfrak
O(R))^{-1}:H^{-1,\alpha}(\mathcal E^R)\to H^{-1,\alpha}(\mathcal
E^R)$,
%\end{equation}
the norm of this operator is bounded by the constant
$1/(1-|\mspace{-2mu}|\mspace{-2mu}|\mathfrak
O(R_0)|\mspace{-2mu}|\mspace{-2mu}|_{R_0})$ uniformly in $R$,
$R>R_0$. We set $\mathcal F'=(I+\mathfrak O(R))^{-1}\mathcal F$.
In the same way as before we construct the approximate solution
$Y$ to the problem~\eqref{e1}--\eqref{e6}, where $\mathcal F$ is
replaced by $\mathcal F'$ and $\mathcal G=0$. Then $\mathcal
A^RY=\{\mathcal F,0\}$, the estimate~\eqref{es9} holds with
$\mathcal G=0$. This %together with the inequality $\|\mathcal
%F';H^{-1,\alpha}(\mathcal E^R)\|\leq \|\mathcal
%F;H^{-1,\alpha}(\mathcal
%E^R)\|/(1-|\mspace{-2mu}|\mspace{-2mu}|\mathfrak
%O(R_0)|\mspace{-2mu}|\mspace{-2mu}|_{R_0})$
proves that in the case $\mathcal G=0$ the problem
\eqref{e1}--\eqref{e6} has a solution $v^{R}\in \tilde
H^{1,\alpha}(\mathcal E^R)$ satisfying the estimate \eqref{est1}.
In the case $\mathcal G\neq 0$ we find an exact solution $v^R$ to
the problem \eqref{e1}--\eqref{e6} in the form $Y_1-Y_2$. Here
$Y_1$ is the approximation solution of the problem
\eqref{e1}--\eqref{e6} with the right hand side $\{\mathcal
F,\mathcal G\}$, and $Y_2$ is the approximation solution of the
problem \eqref{e1}--\eqref{e6} with $\mathcal G=0$ and $\mathcal
F$ replaced by $(I+\mathfrak O(R))^{-1}(\tilde{\mathcal
F}-\mathcal F) $. We have $\mathcal A^R Y_1=\{\tilde {\mathcal
F},\mathcal G\}$ and $\mathcal A^R Y_2=\{\tilde {\mathcal
F}-\mathcal F,0\}$. Now we see that in the case $\mathcal G\neq 0$
the problem \eqref{e1}--\eqref{e6} also has a solution $v^{R}\in
\tilde H^{1,\alpha}(\mathcal E^R)$ satisfying the
estimate~\eqref{est1}. Indeed, by the proved case the
estimate~\eqref{est1} is valid for $v^R=Y_2$, $\mathcal G=0$, and
$\mathcal F$ replaced by $\tilde {\mathcal F}-\mathcal F$. This
together with the estimate~\eqref{es7}, and the
estimate~\eqref{es9} for $Y=Y_1$, leads to~\eqref{est1}. To prove
the uniqueness of the solution $v^{R}$ it suffices to apply the
same argument to the formally adjoint problem.
Proposition~\ref{p5} is proved.

\end{pf}

\begin{thm}\label{t5} Let $T$ be a sufficiently
large positive number. Assume that the potential $q$ and the right
hand side $F$
 satisfy all  the
assumptions of Section~\ref{s2}, the homogeneous  problem
\eqref{op1}--\eqref{op3} has no nontrivial solution in the space
$H^2_{\epsilon}(\mathcal E)$, $\epsilon>0$.  We define the
potential $q_\phi^T$ in \eqref{e2} and the right hand side
$\mathcal F$ of the equations \eqref{e1},\eqref{e2} by the
equalities \eqref{q} and  \eqref{F}.  Let $u$ denote a solution to
the original problem \eqref{op1}--\eqref{op3} with radiation
conditions. Then there exists $R_0>T$ such that for $R>R_0$ a
(unique) variational solution $v^{R}$ to the problem
\eqref{e1}--\eqref{e6}, where $\mathcal G\equiv 0$, converges  to
the solution $u$ in the domain $\mathcal E^T$ in the following
sense
\begin{equation}\label{me}
\|u-v^{R};H^1(\mathcal E^T)\|\leq Ce^{-\gamma R}\|\mathcal F;
H^0_\gamma(\mathcal E)\|, \quad R>R_0,
\end{equation}
where the constant $C$ is independent of $R$ and $\mathcal F$, and
\begin{equation}\label{gamma}
\gamma\in(0,\tau\sin\phi]\cap
(0,\min_n\{\Im(e^{i\phi}\lambda_n^-)\})
\end{equation}
 with the same $\tau>0$ as
in \eqref{tau}.
\end{thm}
\begin{pf}  Due to Proposition~\ref{p3}
a unique solution $u$ to the problem~\eqref{op1}--\eqref{op3} with
radiation conditions
 and a unique variational solution $v\in \lefteqn{\stackrel{\circ}{\phantom{\,\,>}}}H_0^{1,\alpha}(\mathcal E)$ to the scaled
 problem \eqref{lp1}--\eqref{lp5}
are coincident on $\mathcal E^T$. Thus  the estimate~\eqref{me} is
valid if and only if it is valid with $u$ replaced by $v$. The
difference $v-v^R\in \tilde H^{1,\alpha}(\mathcal E^R)$ satisfies
the problem \eqref{e1}--\eqref{e6} with the right hand side
$\mathcal F\equiv 0$ and $\mathcal G=v|_{\Upsilon^R}$. It is clear
that
$$
\|v|_{\Upsilon^R}; H^{1/2,\alpha}(\Upsilon^R)\|\leq e^{-\gamma R}
\|v;
\lefteqn{\stackrel{\circ}{\phantom{\,\,>}}}H^{1,\alpha}_{\gamma}(\mathcal
E)\|.
$$
This together with the first assertion of Proposition~\ref{p3} and
the assumption \eqref{tau} leads to the estimate
$$
\|\mathcal G;H^{1/2,\alpha}(\Upsilon^R)\|\leq C e^{-\gamma
R}\|\mathcal F; H^0_\gamma(\mathcal E)\|
$$
with the same restrictions on $\gamma$ as in \eqref{gamma}. Then
by Proposition~\ref{p5} we have
$$
\|v-v^R;H^1(\mathcal E^R)\|\leq C \|\mathcal
G;H^{1/2,\alpha}(\Upsilon^R) \|\leq C e^{-\gamma R}\|\mathcal F;
H^0_\gamma(\mathcal E)\|,\quad R>R_0.
$$
Theorem~\ref{t5} is proved.
\end{pf}
\begin{rem} It is quite possible that Theorem~\ref{t5} (as well as Propositions~\ref{p3} and \ref{p5}) remains valid
without the assumption on the largeness of the parameter $T$. But
we suppose it all the same because our proof of
Proposition~\ref{p3} is essentially based on this assumption; see
\cite{ref8}.
\end{rem}


\begin{thebibliography}{00}
\bibitem{ref1} J.P. B\'{e}renger, ``A perfectly matched layer for the
absorption of electromagnetic waves'', J. Comp. Phys., vol. 114,
pp. 185--200, 1994.


\bibitem{ref3} M. Lassas, E. Somersalo, ``Analysis of the
PML equations in general convex geometry'', Proc. Roy. Soc.
Edinburgh Sect. A, vol. 131, pp. 1183--1207, 2001.

\bibitem{ref4} T. Hohage, F. Schmidt, L. Zschiedrich, ``Solving
time-harmonic scattering problems based on the pole condition II.
Convergence of the PML method'', SIAM J. Math. Anal., vol. 35 pp.
547--560, 2003.

\bibitem{ref5} E. Becache, A.-S. Bonnet-Ben Dhia, G. Legendre, ``Perfectly matched
layers for the convected Helmholtz equation'', SIAM J. Numer.
Anal., vol. 42, pp. 409--433, 2004

\bibitem{ref5+} E. Becache, A.-S. Bonnet-Ben Dhia, G. Legendre, ``Perfectly
matched layers for time-harmonic acoustics in the presence of a
uniform flow'', SIAM J. Numer. Anal., vol. 44, pp. 1191--1217,
2006.

\bibitem{ref6} V. Maz'ya, S. Nazarov,  B. Plamenevskii, Asymptotic
theory of elliptic boundary value problems in singularly perturbed
domains, Basel, Birkh\"auser (2000).

\bibitem{ref7} V. A. Kozlov,  V. G. Maz'ya, J. Rossmann, Elliptic
boundary value problems in domains with point singularities,
Mathematical Surveys and Monographs, vol. 52 (1997).


\bibitem{ref8} V. Kalvin, ``Radiation conditions and complex
scaling for the stationary Schr\"{o}dinger equation in a periodic
structure'', in preparation

\bibitem{ref9} V. Kalvin, ``Weighted Hardy-Sobolev spaces and
 complex scaling of differential equations with operator coefficients'',
 preprint

\bibitem{ref10}T. Hohage,  ``Laplace domain methods
 for the construction of transparent boundary
  conditions for time-harmonic problems'',
   Mathematical and numerical aspects of wave propagation---WAVES 2003, 148--153, Springer, Berlin, 2003.
% \bibitem{label}
% Text of bibliographic item

% notes:
% \bibitem{label} \note

% subbibitems:
% \begin{subbibitems}{label}
% \bibitem{label1}
% \bibitem{label2}
% If there is a note, it should come last:
% \bibitem{label3} \note
% \end{subbibitems}




\end{thebibliography}
\end{document}